\documentclass[runningheads]{CMSIM}

\usepackage{graphicx} % standard LaTeX graphics tool

\usepackage{amsmath,amssymb}

\usepackage{bm}
\usepackage{enumerate}
\allowdisplaybreaks
\usepackage[usenames]{color} % for colour remarks

\usepackage{bm}
\def\eps{{\varepsilon}}

\newcommand{\fp}[1]{FP$ {\textrm{#1}}$}

\def\S{\mathcal{S}}

\def\eRM{{\mathrm e}}
\def\dRM{{\mathrm d}}

\def\mx{{\bm x}}
\def\mv{{\bm v}}

\def\mk{{\bm k}}

\def\mpp{{\bm p}}
\def\mr{{\bm r}}
\def\mJ{{\bm J}}
\def\mE{{\bm E}}
\def\mB{{\bm B}}

\def\eps{\varepsilon}
\def\boldnabla{{\bm \nabla}}
\allowdisplaybreaks
\allowbreak
\renewcommand{\thefootnote}

\title*{ 
 Passive advection of a vector field: effects of strong compressibility
}

\titlerunning{\it Passive advection of a vector field: effect of strong compressibility }

\author{
 Nikolay~V.~Antonov\inst{1}
 \and 
 Nikolay~M.~Gulitskiy\inst{1}
 \and 
  Maria~M. Kostenko\inst{1}
\and
  Tom\'a\v{s}~Lu\v{c}ivjansk\'y\inst{2,3} 
}

\authorrunning{\it N.~V.~Antonov}

\institute{
Department of Physics,  
Saint-Petersburg State University, 7/9~Universitetskaya nab., St. Petersburg, 199034 Russia
\\
(E-mail: {\tt n.antonov@spbu.ru, n.gulitskiy@spbu.ru, kontramot@mail.ru} )
\and
  Faculty of Sciences, {S}afarik University, Moyzesova 16, 040 01 Ko\v{s}ice, Slovakia \\
  (E-mail: {\tt tomas.lucivjansky@upjs.sk})
   \and 
   Peoples’ Friendship University of Russia (RUDN University), 6 Miklukho-Maklaya
   St, Moscow, 117198, Russian Federation   
}

\begin{document}
\thispagestyle{empty}
\maketitle             
\setlength{\leftskip}{0pt}
\setlength{\headsep}{16pt}
\footnote{\begin{tabular}{p{11.2cm}r}
\small {\it $10^{th}$CHAOS Conference Proceedings, 30 May - 2 June 2017, Barcelona Spain} \\  
 %\small C. H. Skiadas (Ed)\\
   \small \textcopyright {} 2017 ISAST & \includegraphics[scale=0.38]{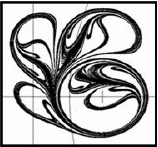}
 \end{tabular}}

%===================================================================================================% 
\begin{abstract}
The field theoretic renormalization group and the operator product expansion are applied
to the stochastic model of a passively advected vector field. 
The advecting velocity field is generated by the stochastic Navier-Stokes
equation with compressibility taken into account. The model is considered in the vicinity
of space dimension $d = 4$ and the perturbation theory is constructed within a double
expansion scheme in $y$ and $\eps=4-d$, where $y$ describes scaling behaviour of the
random force that enters a stochastic equation for the velocity field. We show that the
correlation functions of the passive vector field in the inertial range exhibit anomalous
scaling behaviour. The critical dimensions of tensor composite operators of passive vector
field are calculated in the leading order of $y$, $\eps$ expansion.
%---------------------------------------------------------------------------------------------------%
\end{abstract}
%===================================================================================================%

%---------------------------------------------------------------------------------------------------%
\section{ Introduction }\label{sec:intro}
%---------------------------------------------------------------------------------------------------%
% VSEOBECNY OPIS magnetizmu v astrofyzike
Many natural phenomena in the nature are concerned with hydrodynamic flows. 
Ranging from microscopic up to macroscopic spatial scales 
 fluids can exist in very different states. 
Especially intrigued behaviour is observed for turbulent flows; moreover, 
such flows are rather a rule than an exception \cite{Frisch,Davidson}.
 Despite a vast amount of effort that
 has been put into investigation of turbulence, the problem remains unsolved.
 
 In the astrophysical applications turbulence is quite an ubiquitous phenomenon \cite{Shore,Priest}.  
A very important model is so-called Kazantsev-Kraichnan kinematic model \cite{ant06}.
The basic idea is to assume that a magnetic field is passively advected by
velocity field, but back influence on the velocity field from magnetic field is negligible (for
 a general introduction to magnetohydrodynamic see, e.g., \cite{Moffatt}).
 A genuine model of magnetohydrodynamics (MHD) has to deal with a mutual interplay
 between magnetic field and velocity field. 
 There are many studies \cite{Four82,AVH84} devoted to this problem, mainly
 because it provides a mechanism for a generation of turbulent dynamo \cite{Moffatt}.

% efekt stlacitelnosti 
Especially in an astrophysical context we have to deal with a compressible fluid
rather than incompressible \cite{Shore}. Also in recent years
there has been an activity of compressible MHD turbulence
 \cite{Kim05,Carbone09,Sahra09,Eyink10,Galtier11,Banerjee13,Banerjee16,Hadid17}.
 
In this work, our aim is to look at a compressible turbulence \cite{LL,Sagaut}, motivated by
 the previous studies \cite{Iroshnikov,AG12,J13,Uni3,VectorN} of
the incompressible case and the need for an astrophysical description of
a squishy medium. In case of a compressible medium, we are in fact
 examining conditions for the generation of sound. Any compression generates
acoustic (sound) waves that are transmitted through the medium and serve
as the prime source for dissipation. So the problem of the energy spectrum
(and dissipation rate) of a compressible fluid is essentially one of stochastic
acoustics.

% RG pristup
%---------------------------------------------------------------------------------------------------%
The investigation of such behaviour as anomalous scaling requires a lot of thorough, even
meticulous, analysis to be carried out. The phenomenon manifests itself
in a singular (arguably, power-like) behaviour of some statistical quantities (correlation functions, 
structure functions, etc.) in the
inertial-convective range in the fully developed turbulence regime~\cite{Frisch,Davidson,Falkovich2001}.

A quantitative parameter that describes ``strength'' of turbulent motion is so-called 
Reynolds number $\mathrm{Re}$ {which
represents} a ratio between inertial and dissipative forces. For high enough values of 
$\mathrm{Re} \gg 1$
 inertial interval is exhibited in which just transfer of kinetic energy from outer $L$ (input)
  to microscopic $l$ (dissipative) scales take place. 
%-------------------------------------------------------------------------------%

A very useful and computationally effective approach to the problems with many interacting degrees of 
freedom on different scales is the field-theoretic 
renormalization group (RG) approach which can be subsequently accompanied by the operator product
expansion (OPE); see the monographs~\cite{Vasiliev,Zinn,turbo,Tauber}. 
One of the greatest challenges is an investigation of the Navier-Stokes equation for a compressible 
fluid, and, in particular, 
 a passive scalar field advection by this velocity ensemble. The first relevant discussion and analysis 
 of passive advection emerged a few 
decades ago for the Kraichnan's velocity ensemble~\cite{Kraich1,GK,RG}. Further studies developed
 its more realistic 
generalizations~\cite{AG12,J13,Uni3,VectorN,JJ15,V96-mod,amodel,HHL}. The RG+OPE technique 
 was also applied to more complicated models, in 
particular, to the 
compressible case~\cite{NSpass,Ant04,AGM,JJR16,VM,tracer2,tracer3,ANU97,AK14,AK15,AK2,LM,St,MTY}.
%-------------------------------------------------------------------------------%
%-------------------------------------------------------------------------------%

The paper is a continuation of our previous works~\cite{AGKL16,AGKL17_epj,AGKL17_pre} and is
organized as follows. In the 
introductory Sec.~\ref{sec:model} 
we give a brief overview of the model and we reformulate stochastic equations
into field-theoretical language.
 Sec.~\ref{sec:RG} is devoted to the renormalization group analysis.
In Sec.~\ref{sec:scaling} we present
 the fixed points' structure, describe possible scaling regimes and calculate
  critical dimensions. 
  In Sec.~\ref{sec:ope} OPE is applied to the equal-time structure functions constructed of the 
 vector fields; the anomalous exponents are 
calculated. The concluding Sec.~\ref{sec:conclusion} is devoted to a brief discussion.
%-------------------------------------------------------------------------------%

%---------------------------------------------------------------------------------------------------%
\section{ Model }\label{sec:model}
%---------------------------------------------------------------------------------------------------%

Let us start with a brief discussion of a model for compressible velocity fluctuations. 
The dynamics of a compressible fluid is governed by the Navier-Stokes equation~\cite{LL}:
\begin{equation}
  \rho\nabla_{t} v_{i} = \nu_{0} [\delta_{ik}\partial^{2}-
  \partial_{i}\partial_{k}] v_{k}
  + \mu_0 \partial_{i}\partial_{k} v_{k} - \partial_{i} p + f^v_{i},
  \label{eq:NS}
\end{equation}
where the operator $\nabla_t$ denotes an expression $\nabla_{t} = \partial_{t} + v_{k} \partial_{k}$, 
also known as a Lagrangian (or convective) derivative. Further,  
$\rho=\rho(t,\mx)$ is a fluid density field, $v_i=v_i(t,\mx)$ is the velocity field,
$\partial_{t} =
\partial /\partial t$, $\partial_{i} = \partial /\partial x_{i}$, 
$\partial^{2} =\partial_{i}\partial_{i}$ is the Laplace operator, $p=p(t,\mx)$ is the pressure field,  
and $f^v_i$ is the external force, which is specified later. 
In what follows
 we employ a condensed notation in which we write  $x=(t,{\bm x})$, where a spatial
  variable $\mx$ equals $(x_1,x_2,\ldots,x_d)$ with $d$ being a dimensionality of space.
Two parameters $\nu_{0}$ and $\mu_{0}$ are two
 viscosity coefficients~\cite{LL}. Summations over repeated vector indices (Einstein summation
 convention) are always implied in this work.

 Let us note two important remarks regarding the interpretation of Eq.~(\ref{eq:NS}). First, this equation should 
 be regarded as an equation only for a fluctuating part of the total velocity field. In other words,
 it is implicitly assumed that the mean (regular) part of the velocity field has been subtracted 
 \cite{Frisch,Davidson}.
 Second, the random force $f_i^v$ not only mimics an input of energy, but to some extent
 it is responsible for neglected
 interactions between fluctuating part of the velocity field and the mean part \cite{turbo}. 
 In reality the latter interactions are always present and their mutual interplay
 generates turbulence \cite{Davidson}. In a sense, stochastic theory of turbulence
 is similar to a fluctuation theory for critical phenomena \cite{Vasiliev,papo}.
 
To finalize the theoretical description of velocity fluctuations, Eq.~(\ref{eq:NS}) 
has to be augmented by additional two relations. They are a continuity equation and
 a certain thermodynamic relation \cite{LL}.
 The former can be written in the form
\begin{equation}
  \partial_{t} \rho  + \partial_{i} (\rho v_{i})   = 0
  \label{eq:CE}
\end{equation}
and the latter {we choose as}
\begin{equation}
  \delta p = c_0^2 \delta\rho,
  \label{eq:SE}
\end{equation}
where $\delta p$ and $\delta \rho$ describe deviations from the equilibrium values of pressure field
 and density field, respectively.

Viscous terms describe dissipative processes in the system and are especially important
 at small spatial scales. Without a continuous input of energy
turbulent processes would eventually die out and {the flow become} regular.
 There are various possibilities for modelling of energy input \cite{turbo}. For translationally
 invariant theories it is convenient to specify properties of the random force $f_i$ in 
 frequency-momentum representation
\begin{equation}
  \langle f_i(t,{\mx}) f_j(t',{\mx}') = \frac{\delta(t-t')}{(2\pi)^d} \int_{k>m} \dRM^d k \mbox{ }
  D^v_{ij}({\mk})
  \eRM^{i{\mk}\cdot({\mx-\mx'})},
\label{eq:ff}
\end{equation}
where the delta function ensures Galilean invariance of the model. The integral
is  infrared~(IR) regularized with a parameter $m\sim L_v^{-1}$, where 
$L_v$ denotes outer scale, i.e., scale of the biggest turbulent eddies. More
details can be found in the literature \cite{turbo,JETP}. 
 The kernel function ${D}^v_{ij}({\bm k})$ is now chosen in the following form
\begin{equation}  
  D_{ij}^v({\bm k})=g_{10} \nu_0^3 k^{4-d-y} \biggl\{
  P_{ij}({\bm k}) + \alpha Q_{ij}({\bm k})
  \biggl\} 
  + g_{20} \nu_0^3 \delta_{ij}
  \label{eq:correl2}
\end{equation}
that consists of two terms. The term proportional to the charge $g_{10}$ is non-local and
ensures a steady input of energy into the system from outer scales. In what follows we employ
 the RG approach. The value of the scaling exponent $y$ describes a deviation from a
logarithmic behaviour.
In the stochastic theory of turbulence the main interest is in the limit behaviour
 $y\rightarrow 4$ that yields an ideal pumping from infinite spatial scales \cite{turbo}.
 The projection operators $P_{ij}$ and $Q_{ij}$
 in the momentum space read
\begin{equation}
   P_{ij} (\mk) = \delta_{ij} - \frac{k_i k_j}{k^2}, \quad Q_{ij} = \frac{k_i k_j}{k^2}
   \label{eq:project}
\end{equation}
and correspond to the transversal and longitudinal projector, respectively, $k=|\mk|$ is the 
wave number.
The local term proportional to $g_{20}$ in (\ref{eq:project})
is not dictated by the physical considerations, but rather by a proper 
renormalization treatment~\cite{AGKL17_pre}. Let us briefly describe this subtle point.
 An important difference of the present study with the traditional approaches 
\cite{ant06,turbo} is a special role of the space dimension~$d=4$. 
Usually the spatial dimension $d$ plays a passive role and
  is considered only as an independent parameter. However, Honkonen and Nalimov \cite{HN96}
  showed that in the vicinity of space dimension {$d=2$} additional divergences appear in the model
  of the incompressible Navier-Stokes ensemble and
  these divergences have to be properly taken into account. Their procedure also results
  into improved perturbation expansion \cite{AHKV03,AHH10}.
    As we see in the next section a similar situation occurs for the model (\ref{eq:NS})
   in the vicinity of space dimension $d=4$. In this case an additional divergence
   appears in the 1-irreducible Green function $\left\langle v'v'\right\rangle_{\text{1-ir}}$. 
   This feature allows us to employ a double expansion scheme, in which the 
formal expansion parameters are $y$, which describes the scaling behaviour of a
random force, and $\eps=4-d$, i.e., a deviation from the space dimension $d=4$ \cite{HHL,HN96}. 

%--------------------------------------------------------------------------------------------------------
%		INTRO OF MAGNETIC FIELD
%--------------------------------------------------------------------------------------------------------
The inclusion of magnetic field in Kazantsev-Kraichnan model follows a simple
physical reasoning. The first important assumption is that conditions
for a so-called MHD limit are met. Broadly speaking, this corresponds
to a dense limit in which the charge and bulk densities are obtained rather from the fluid equations 
and not from the Boltzmann equation \cite{Shore}.
 The second assumption is that the current is connected with the electromagnetic fields via
\begin{equation}
   {\mJ} = \sigma({\mE} + \mv\times{\mB}), 
\end{equation}
where $\mE$ is an electric field, $\mB$ is a magnetic field, and $\sigma$ is the conductivity of a medium.
Neglecting Maxwell displacement current one can finally derive 
following 
\begin{equation}
  \partial_t \theta_i + \partial_k(v_k\theta_i - v_i \theta_k) = \kappa_0 \partial^2 \theta_i + f^\theta_i,
  \label{eq:eom_mag}
\end{equation}
where $\kappa_0$ is the magnetic diffusion
 coefficient. For a detailed exposition we recommend textbooks \cite{Shore,Moffatt}.
 Note that in stochastic approach to MHD, Eq. (\ref{eq:eom_mag}) should be understood as
 an equation for the fluctuating part $\theta_i=\theta_i(x)$ of the total magnetic 
 field \cite{Four82,AVH84,ALM00}.

 Random force $f_i^\theta\equiv f_i^\theta(x)$ is again assumed to be a Gaussian variable
 with zero mean and given covariance,
\begin{equation}
  \langle f_i^\theta(x) f_j^\theta (x') \rangle = \delta(t-t')\, C_{ij}({\mr}/L_\theta), \quad
  {\mr}= {\mx} - {\mx}',
  \label{eq:noise}
\end{equation}
where $C_{ij}({\bf r}/L_\theta)$ is a certain
function finite at limit $({\mr}/L_\theta)\to 0$ and rapidly decaying for
$({\mr}/L_\theta)\to\infty$.
An additional condition for the magnetic
field arises (namely, transversality condition $\partial_i \theta_i = 0$), which makes the terms
$\partial_{k}(v_{i}\theta_k)$ and $(\theta_k \partial_{k})v_i $ equal.
 Let us mention that $L_\theta$ is an integral scale related to the 
  stirring of magnetic field, and $C_{ij}$ is a
function finite in the limit $L_\theta\rightarrow\infty$. A 
 detailed form of the function $C_{ij}$
is not relevant. The only condition that must be satisfied is that $C_{ij}$ decreases rapidly for
 $r\gg L_\theta$.  In a physically more realistic formulation,
the noise might be replaced, e.g., by the term $(\mB\cdot\boldnabla)$, where 
$\mB$ is a constant large-scale magnetic
field (see, e.g., \cite{AK15,ALM00}). It is
worth to mention that we always assume that inequality $L_\theta \gg L_v$ holds.

In more realistic scenarios there should be an additional Lorentz term in Eq. (\ref{eq:NS}), which
would correspond to the active advection of magnetic field. This would require presence of the
Lorentz term
 \begin{equation*}
   \mv\times\mB \sim {\bm J} \sim (\boldnabla\times \mB)\times \mB. 
 \end{equation*}
 As has been pointed out, in this work we restrict ourselves
to a kinematic approximation in which such term is not included in the model.

Our main theoretical tool is the renormalization group theory. Its proper application requires
 a proof of a renormalizability of the model, i.e., a proof that only a finite number 
 of  divergent structures exists in a diagrammatic expansion \cite{Zinn,Amit}.
 As was shown in \cite{VN96}, this requirement can be accomplished 
 by the following procedure:
 first the  stochastic equation~(\ref{eq:NS}) is divided by density field $\rho$, 
 then fluctuations in viscous terms are neglected, and finally.
 using the expressions~(\ref{eq:CE}) and~(\ref{eq:SE})
the problem is formulated into a system of two coupled equations
\begin{align}
  \nabla_{t} v_{i} & = 
  \nu_{0} [\delta_{ik}\partial^{2}-\partial_{i}\partial_{k}]
  v_{k}\! +\! \mu_0 \partial_{i}\partial_{k} v_{k} -\!
  \partial_{i} \phi\! +\! f_{i},
  \label{eq:ANU} \\
  \nabla_{t} \phi & =  -c_{0}^{2} \partial_{i}v_{i},
  \label{eq:ANU1}
\end{align}
where a new field $\phi=\phi(x)$ has been introduced and it is related to the density fluctuations via the 
relation $\phi = c_0^2 \ln (\rho/\overline{\rho})$ \cite{AGKL17_pre,VN96}.
A parameter $c_0$ denotes the adiabatic speed of sound, $\overline{\rho}$ is the mean value of $\rho$, and 
$f_{i}=f_{i}(x)$ is the external force normalized per unit mass.

According to the general theorem~\cite{Vasiliev,Zinn}, the stochastic problem given by Eqs.
 (\ref{eq:eom_mag}),(\ref{eq:ANU}), and (\ref{eq:ANU1}),
is tantamount to the field theoretic
model with a doubled set of fields $\Phi=\left\{v_{i}, v_{i}',\phi, \phi'\right\}$ and De Dominicis-Janssen action 
functional. The latter can be written
in a compact form as a sum of two terms 
\begin{align}
  \S_\text{total} [\Phi] & = \S_\text{vel}[\Phi] + \S_\text{mag}[\Phi], 
  \label{eq:full_action} 
\end{align}
where the first term describes a velocity part
\begin{align}
  \S_\text{vel}[\Phi] & = \frac{v_i' {D}_{ij}^v v_j'}{2} 
  +v_i' \biggl[
  -\nabla_t v_i + \nu_0(\delta_{ij}\partial^2 - \partial_i \partial_j)v_j
  +u_0 \nu_0 \partial_i \partial_j v_j - \partial_i \phi
  \biggl] \nonumber \\  &
  +\phi'[-\nabla_t \phi  + v_0 \nu_0 \partial^2 \phi - c_0^2 (\partial_i v_i)].
  \label{eq:vel_action} 
\end{align}  
Here, $ {D}^v_{ij}$ is the correlation function~(\ref{eq:correl2}). Note that we have introduced 
a new dimensionless parameter
$u_0=\mu_0/\nu_0>0$ and a new term
$v_0 \nu_{0} \phi' \partial^{2}\phi$ with another positive
dimensionless parameter $v_0$, which is needed  to ensure 
multiplicative renormalizability. 

  The second term in Eq.(\ref{eq:full_action}) reads
\begin{equation}  
  \S_\text{mag}[\Phi]  = 
   \frac{1}{2} \theta_i' D^\theta_{ij} \theta_j' + \theta_k'[ 
  -\partial_t \theta_k -(v_i\partial_i)\theta_k + (\theta_i\partial_i)v_k + \nu_0 w_0 \partial^2 \theta_k
  ],
  \label{eq:mag_action} 
\end{equation}
where we have introduced another dimensionless parameter $w_0$ via $\kappa_0=\nu_0 w_0$.
Also we have employed a condensed notation, in which integrals over the spatial variable 
${\mx}$ and the time variable $t$, as well as summation over repeated indices, are implicitly assumed, 
for instance
\begin{align}  
  {\phi'}\partial_t{\phi} & =\int\! \dRM t \!\int \dRM^d{x}\, \phi'(t,\mx)\partial_t\phi(t,\mx), \nonumber \\
  v'_iD_{ik}{v'}_k
  & = \int\! \dRM t \!\int\! \dRM^d x \!\! \int\! \dRM^d x' 
  \,v_i(t,{\mx})D^v_{ik}({\mx}-{\mx'})v_k(t,{\mx'}). 
  \label{eq:quadlocal2}
\end{align}

In a functional formulation various stochastic
quantities (correlation and structure functions) are calculated as path
integrals with weight functional 
\begin{equation*}
  \exp (S_\text{total}[\Phi]). 
\end{equation*}
 The main benefits of such approach are transparence in a perturbation theory and
 the powerful methods
of the quantum field theory, such as Feynman diagrammatic technique and 
renormalization group procedure \cite{Zinn,turbo,Tauber}.
 
\begin{figure}
  \centerline{
    \includegraphics[width=0.9\textwidth]{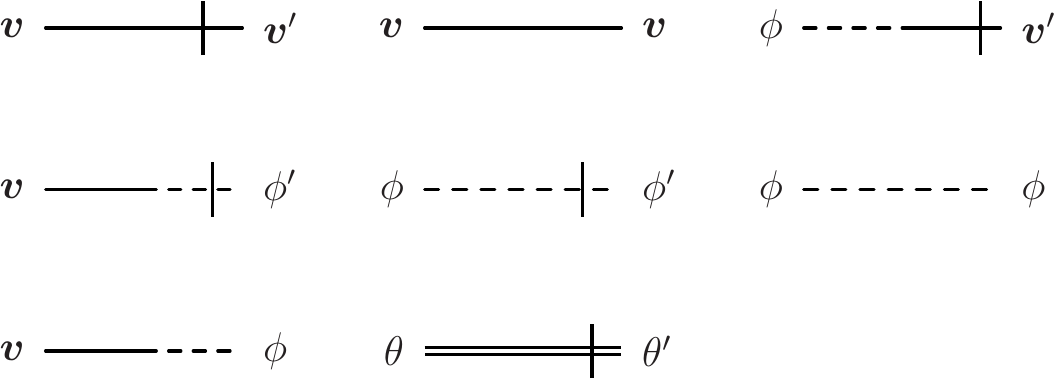}}
  \caption{Graphical representation of all propagators of the model given by the quadratic part of the 
  action (\ref{eq:full_action}). }
  \label{fig:prop}
\end{figure}

\begin{figure}
  \centerline{
    \includegraphics[width=\textwidth]{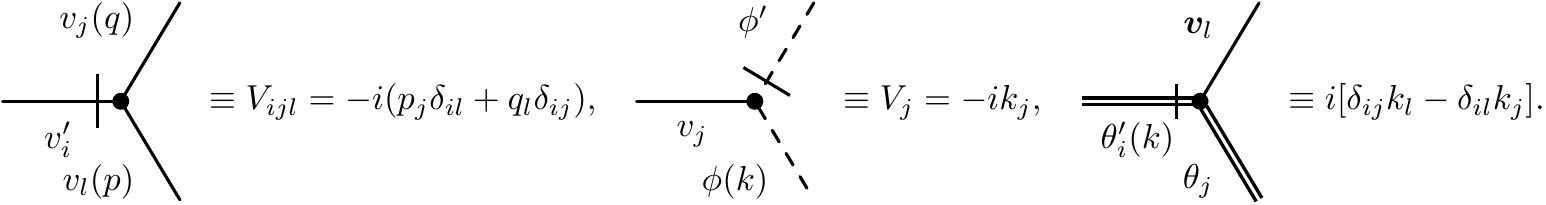}}
  \caption{Graphical representation of all interaction vertices of the model given by the 
  nonlinear part of the action (\ref{eq:full_action}).}
  \label{fig:vert}
\end{figure}
%---------------------------------------------------------------------------------------------------%
\section{ Renormalization group analysis }\label{sec:RG}
%---------------------------------------------------------------------------------------------------%

Ultraviolet renormalizability reveals itself in a presence divergences 
in Feynman graphs, which are constructed according to simple laws \cite{Vasiliev,Tauber} using
 a graphical notation from Figs. \ref{fig:prop} and \ref{fig:vert}.
From a practical point of view,
 an analysis of the 1-particle irreducible Green functions, later referred to as 1-irreducible Green functions
 following the notation in~\cite{Vasiliev}, is of utmost importance. 
In the case of dynamical models \cite{Vasiliev,Tauber}
two independent scales have to be introduced: the time scale $T$ and the length scale $L$. Thus
the canonical dimension of any quantity $F$ (a field or a parameter) is
described by two numbers, the
frequency dimension $d_{F}^{\omega}$ and the momentum dimension $d_{F}^{k}$,
defined such that
\begin{equation}
  d_k^k =-d_{ x}^k=1,\quad d_k^{\omega} =d_{x}^{\omega }=0,\quad
  d_{\omega }^{\omega }=-d_t^{\omega }=1,\quad d_{\omega }^k=d_t^k=0,
  \label{eq:def_dim}
\end{equation}
% \begin{align}
%   d_k^k&=-d_{ x}^k=1, &d_k^{\omega}& =d_{x}^{\omega }=0,\nonumber\\
%   d_{\omega }^{\omega }&=-d_t^{\omega }=1, &d_{\omega }^k&=d_t^k=0,
%   \label{eq:def_dim}
% \end{align}
and the given quantity then scales as
\begin{eqnarray}
[F] \sim [T]^{-d_{F}^{\omega}} [L]^{-d_{F}^{k}}.
\label{eq:canon}
\end{eqnarray}
The remaining dimensions can be found from the requirement that each term of the
action functional (\ref{eq:full_action}) be dimensionless, with respect to both the momentum and the
frequency dimensions separately.

Based on $d_F^k$ and $d_F^\omega$  the total canonical dimension
$d_F=d_F^k+2d_F^\omega$ can be introduced, which in
the renormalization theory of dynamic models plays the
same role as the conventional (momentum) dimension does in
static problems \cite{Vasiliev}. 
 Setting $\omega \sim k^{2}$ ensures that all the viscosity and diffusion coefficients in the model are dimensionless. 
Another option is to set the speed of sound $c_{0}$
dimensionless and  consequently obtain that $\omega \sim k$, i.e., $d_{F}=d_{F}^{k}+d_{F}^{\omega}$.
This variant would mean that
we are interested in the asymptotic behaviour of the Green functions
as $\omega \sim k \to 0$, in other words, in sound modes in turbulent medium. Even though this problem
is very interesting itself, it is not yet accessible {for} the
RG treatment, so we do not discuss it here.
The choice $\omega \sim k^{2} \to 0$ is the same as in the models 
of incompressible fluid, where
it is the only possibility because the speed of sound is infinite.
A similar alternative in dispersion laws exists, for example, within the so-called model H of equilibrium
dynamical critical behaviour, see~\cite{Vasiliev,Tauber}.

The canonical dimensions for the velocity part of the model~(\ref{eq:vel_action}) are listed in
Tab.~\ref{tab:vel}, whereas parameters of the magnetic part are given in Tab.~\ref{tab:mag}. From 
Tabs.~\ref{tab:vel} and \ref{tab:mag} it follows that the model
is logarithmic (the coupling constants $g_{10} \sim [L]^{-y}$
and $g_{20} \sim [L]^{-\varepsilon}$
become dimensionless) at $y=\varepsilon=0$.
In this work we use the minimal subtraction (MS) scheme for the calculation
of renormalization constants. In this scheme the UV divergences in the
Green functions manifest
themselves as poles in $y$, $\varepsilon$ and their linear combinations. Here, in accordance
with critical phenomena we retain the notation $\varepsilon=4-d$ .

\begin{table*}
\centering
\caption{Canonical dimensions of the fields and parameters entering velocity part of the 
total action (\ref{eq:vel_action}).}
\label{tab:vel}
  \begin{tabular}{c|c|c|c|c|c|c|c|c|c|c}
    $F$ & $ v_i'$ & $ v_i$ & $\phi'$ & $\phi$  &
    $m$, $\mu$, $\Lambda$ & $\nu_0$, $\nu$ & $c_{0}$, $c$ &
    $g_{10}$ & $g_{20}$ & $u_{0}$, $v_{0}$ $w_{0}$, $u$, $v$, $g_1$, $g_2$, $\alpha$  \\
    \hline
    $d_{F}^{k}$ & $d+1$ & $-1$ & $d+2$ & $-2$  
    & 1 &  $-2$ & $-1$ & $y$ & $4-d$ & 0 \\
    $d_{F}^{\omega}$ & $-1$ & 1 & $-2$ & 2 & 
    0 & 1 & 1 & 0 & 0 & 0\\
    $d_{F}$ & $d-1$ & 1 & $d-2$ & 2 &  1 & 0 & 1 & $y$ & $4-d$ & 0 \\
  \end{tabular}
\end{table*}

\begin{table*}
\centering
\caption{Canonical dimensions of the fields and parameters entering magnetic part of the 
total action (\ref{eq:mag_action}).}
\label{tab:mag}
  \begin{tabular}{c|c|c|c|c|c|c|c|c|c|c}
    $F$ & $ \theta' $ &  $\theta$ &
     $\kappa$, $\kappa_0$  & 
       $w_{0}$, $w$  \\
    \hline
    $d_{F}^{k}$  & $d$  & 0
    &   $-2$  & 0 \\
    $d_{F}^{\omega}$ & $1/2$ & $-1/2$ &
     1  & 0\\
    $d_{F}$ & $d+1$ & $-1$ & 1  & 0 \\
  \end{tabular}
\end{table*}

The total canonical dimension of any 1-irreducible Green function $\Gamma$ is given by the relation
\begin{eqnarray}
\delta_{\Gamma} = d+2 - \sum_{\Phi} N_{\Phi} d_{\Phi},
\label{index}
\end{eqnarray}
where $N_{\Phi}$ is the number of the given type of field {entering the function}
$\Gamma$, $d_{\Phi}$ is the corresponding total canonical dimension of field $\Phi$, and
the summation runs
over all types of the fields $\Phi$ in function $\Gamma$~\cite{Vasiliev,Zinn,Tauber}.  

Superficial UV divergences whose removal requires counterterms can be present only in
those functions $\Gamma$ 
for which
the formal index of divergence $\delta_{\Gamma}$ is a non-negative integer.
A dimensional analysis should be augmented by the several additional considerations.
 They are clearly stated in the previous work \cite{AK15,AGKL17_pre}. Therefore, we do not repeat them here
and continue with a simple conclusion that model with the action (\ref{eq:full_action}) is
renormalizable. The only graphs that are needed to be calculated are
two-point Green functions. For a velocity part, the following graphs have to be analyzed
\begin{align} 
  &  \raisebox{-1.ex}{ \includegraphics[width=3truecm]{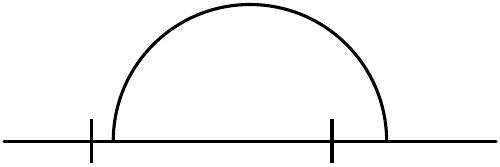}} 
     \raisebox{-1ex}{ \includegraphics[width=3truecm]{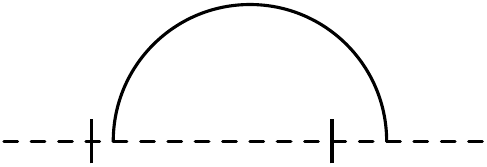}}
     \raisebox{-1ex}{ \includegraphics[width=2.75truecm]{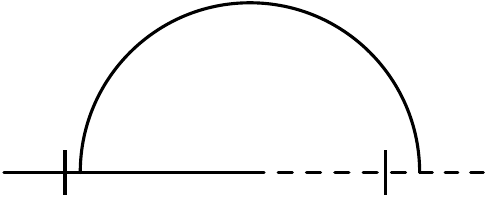}}
     \raisebox{-1ex}{ \includegraphics[width=3truecm]{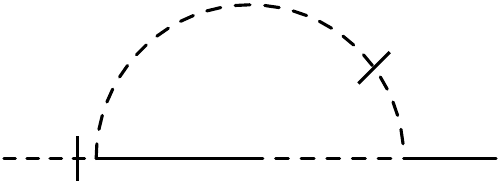}} \nonumber \\
  &  \raisebox{-1ex}{ \includegraphics[width=2.75truecm]{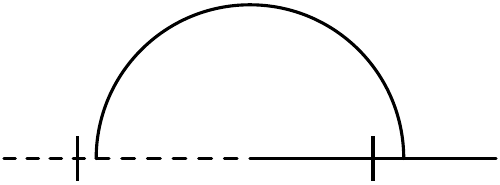}}      
     \raisebox{-1ex}{ \includegraphics[width=2.75truecm]{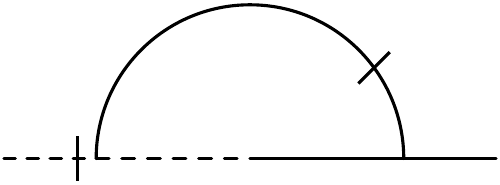}}
     \raisebox{-1.0ex}{ \includegraphics[width=3truecm]{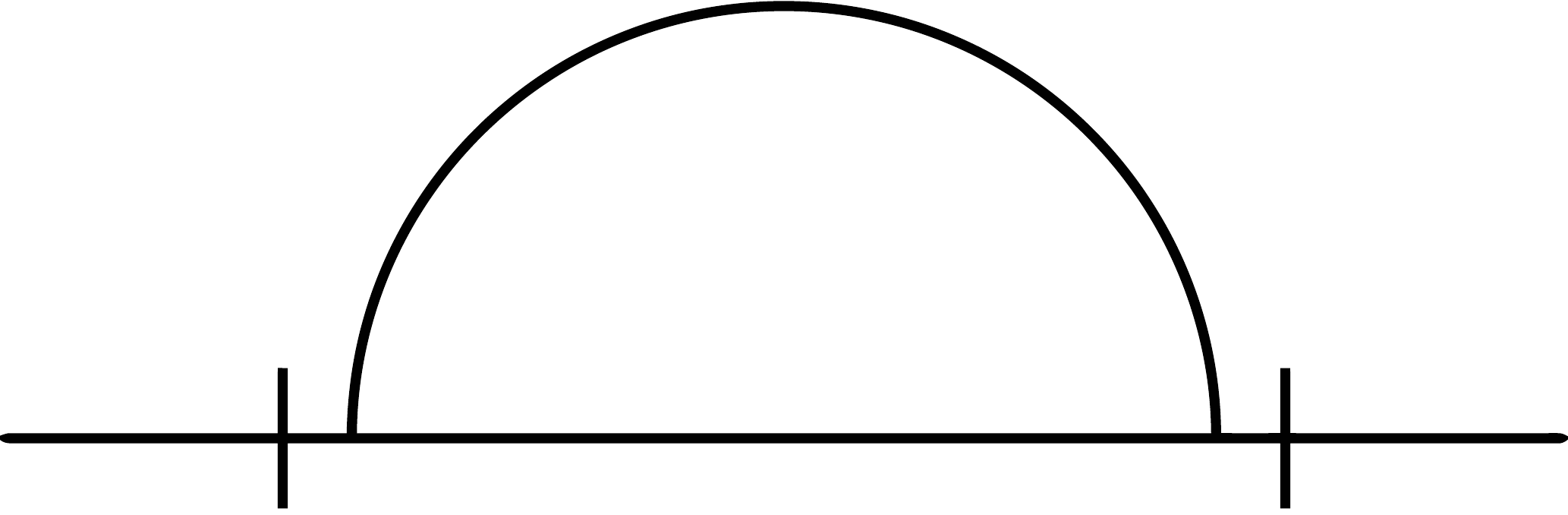}}     
     \label{eq:vel_grafy}
\end{align}
and for a magnetic part we have one Feynman diagram
\begin{align} 
  \raisebox{-1.ex}{ \includegraphics[width=3.5truecm]{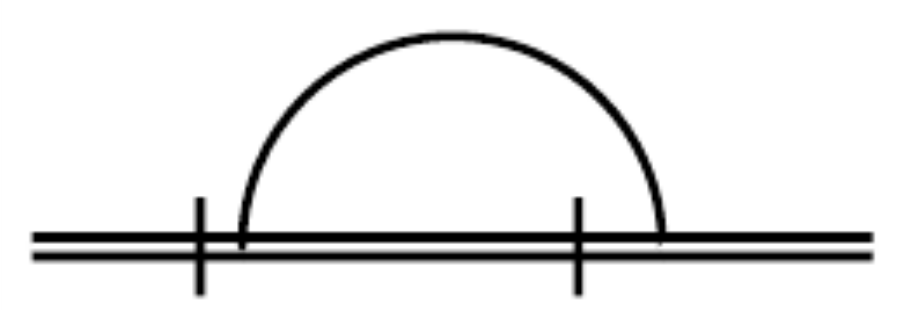}}. 
  \label{eq:mag_grafy}
  %  \raisebox{-5.5ex}{ \includegraphics[width=3.truecm]{triple.pdf}}.     
\end{align}
The remaining diagrams are either UV finite or the Galilean invariance prohibits
their presence. Because the calculation of the divergent parts of Feynman diagrams is rather straightforward
and proceeds in the usual fashion \cite{Vasiliev,Zinn,Tauber,Amit}, we refrain from
mentioning here all the technical details. For the latter, we recommend an interested reader to consult
our previous works \cite{AK15,AGKL16,AGKL17_epj,AGKL17_pre}. In what follows, we focus
on important results that follows for the MHD model (\ref{eq:mag_action}).

Here, we just provide a result of the diagram {$D$ shown in Eq.~(\ref{eq:mag_grafy})}

\begin{align}
   D=\frac{ \overline{S_d} } {2d}p^2 P_{12}(\mpp)\nu 
   &\biggl\{
   \frac{1-d}{1+w}\biggl[ \frac{g_1}{y}\biggl(\frac{\mu}{m}\biggl)^y +
   \frac{g_2}{\eps}\biggl(\frac{\mu}{m}\biggl)^\eps 
   \biggl]\nonumber\\
   &{-
   \frac{u-w}{u(u+w)} 
   \biggl[ \frac{\alpha g_1}{y}\biggl(\frac{\mu}{m}\biggl)^y +
   \frac{g_2}{\eps}\biggl(\frac{\mu}{m}\biggl)^\eps 
   \biggl]   
  \biggl\} }
  \label{eq:mag_feynman1}
\end{align}
where $\overline{S_d}=S_d/(2\pi)^d$ with $S_d = 2\pi^{d/2}/\Gamma(d/2)$ is the
surface area of the unit sphere in the $d-$dimensional space and $\Gamma(x)$ is
Euler's Gamma function.
The expression (\ref{eq:mag_feynman1})
 differs from the result obtained in \cite{AK15} by the presence of terms containing 
the charge $g_2$.

Further, from (\ref{eq:mag_feynman1}) we directly derive renormalization constant $Z_\kappa$ {[where $\kappa=\nu w$, see Eq.~(\ref{eq:mag_action})]}
\begin{equation}
  Z_\kappa = 1 - \frac{g_1}{2dwy}\biggl[ 
  \frac{d-1}{1+w} + \frac{\alpha(u-w)}{u(u+w)^2}
  \biggl] - \frac{g_2}{2dwy} \biggl[ 
  \frac{d-1}{1+w} + \frac{(u-w)}{u(u+w)^2}
  \biggl]
\end{equation}
and the corresponding anomalous dimension 
\begin{equation}
  \gamma_\kappa = \frac{g_1}{2dwy}\biggl[ 
  \frac{d-1}{1+w} + \frac{\alpha(u-w)}{u(u+w)^2}
  \biggl] + \frac{g_2}{2dwy} \biggl[ 
  \frac{d-1}{1+w} + \frac{(u-w)}{u(u+w)^2}
  \biggl].
\end{equation}

%---------------------------------------------------------------------------------------------------%
\section{ Scaling regimes  }\label{sec:scaling}
%---------------------------------------------------------------------------------------------------%

The relation between the initial and renormalized action functionals $\S(\Phi,e_{0})= \S^{R}(Z_\Phi\Phi,e,\mu)$ (where $e_{0}$
is the complete set of bare parameters and $e$ is the set of their renormalized
counterparts) yields the fundamental RG differential equation:
\begin{equation}
\biggl\{ {\cal D}_{RG} + N_{\phi}\gamma_{\phi} +
N_{\phi'}\gamma_{\phi'} \biggr\} \,G^{R}(e,\mu,\dots) = 0,
\label{RG1}
\end{equation}
where $G =\langle \Phi\cdots\Phi\rangle$ is a correlation function of the fields~$\Phi$;
$N_{\phi}$ and $N_{\phi'}$ are the counts of normalization-requiring fields $\phi$ and $\phi'$, respectively, which are the inputs to
$G$; 
the ellipsis in expression~\eqref{RG1} stands for the other arguments of $G$ (spatial
and time variables, etc.).
${\cal D}_{RG}$ is the operation $\widetilde{\cal D}_{\mu}$
expressed in the renormalized variables and
$\widetilde{\cal D}_{\mu}$
is the differential operation $\mu\partial_{\mu}$ for fixed
$e_{0}$. For the present model it takes the form
\begin{equation}
{\cal D}_{RG}= {\cal D}_{\mu} + \beta_{g_1}\partial_{g_1} + \beta_{g_2}\partial_{g_2} +
\beta_{u}\partial_{u} + \beta_{v}\partial_{v}
- \gamma_{\nu}{\cal D}_{\nu}- \gamma_{c}{\cal D}_{c} .
\label{RG2}
\end{equation}
Here, we have denoted ${\cal D}_{x} \equiv x\partial_{x}$ for any variable $x$.
The anomalous dimension $\gamma_{F}$ of some quantity $F$
(a field or a parameter) is defined as
\begin{equation}
\gamma_{F}= Z_{F}^{-1} \widetilde{\cal D}_{\mu} Z_{F} =
\widetilde{\cal D}_{\mu} \ln Z_F ,
\label{RGF1}
\end{equation}
and the $\beta$ functions for the four dimensionless coupling
constants $g_1$, $g_2$, $u$ and $v$,
which express the flows of parameters under
the RG transformation, are $\beta_{g} = \widetilde{\cal D}_{\mu} g$. 
This yields

\begin{align}
  \beta_{g_1} & =  g_1\,(-y-\gamma_{g_1}),
  &\beta_{g_2} & =  g_2\,(-\varepsilon-\gamma_{g_2}),
  &\beta_{u} & =  -u\gamma_{u},\nonumber \\ 
  \beta_{v} & = -v\gamma_{v},
  &\beta_w & = w(\gamma_\nu - \gamma_\kappa).
  \label{eq:all_beta}
\end{align}
The last term follows from the introduced definition of the charge $w$ in Eq.(\ref{eq:mag_action}).
 Based on the analysis of the RG equation~\eqref{RG1} it follows that 
the large scale behaviour with respect to spatial and time scales is
governed by the IR attractive (``stable'') fixed points $g^*\equiv\{g_1^*,g_2^*,u^*,v^*\}$,
 {whose} coordinates are found from the conditions~\cite{Vasiliev,Zinn,Amit}:
\begin{align}
  &\beta_{g_1} (g^{*}) = \beta_{g_2} (g^{*})= \beta_{u}
  (g^{*}) = \beta_{v} (g^{*}) = 0.
  \label{eq:gen_beta}
\end{align}
 Let us consider a set of invariant couplings $\overline{g}_i = \overline{g}_i(s,g)$
with the initial data $\overline{g}_i|_{s=1} = g_i$. Here, $s=k/\mu$ 
and IR asymptotic behaviour (i.e., behaviour at large distances) corresponds
to the limit $s\rightarrow 0$. An evolution of invariant couplings is described by
the set of flow equations
\begin{equation}
  \mathcal{D}_s \overline{g}_i = \beta_i(\overline{g}_j),
  \label{eq:invariant_chrg}
\end{equation}
whose solution as $s\to0$ behaves approximately like
\begin{equation}
  \overline{g}_i(s,g^*) \cong g^*+const\times s^{\omega_i},
  \label{Asym}
\end{equation}
where $\left\{\omega_i\right\}$ is the set of eigenvalues of the matrix 
\begin{equation}
\Omega_{ij}=\partial\beta_{i}/\partial g_{j}|_{g=g_{*}}.
\label{Omega}
\end{equation}
The existence of IR {attractive} solutions of the RG equations leads
to the existence of the scaling behaviour of Green functions. 
From~\eqref{Asym} it follows that the type of the fixed point is determined by the matrix~\eqref{Omega}:
for the IR {attractive} fixed points the matrix $\Omega$ has to be positive definite.

 The character
of the IR behaviour depends on the mutual relation between $y$ and $\varepsilon$~-- two formally small
quantities which were introduced 
in the correlator of the random force in the Navier-Stokes equation. In practical calculations
they constitute parameters into which universal quantities are expanded. This is done in a similar
fashion as calculation of critical exponents in $\phi^4$ theory, see~\cite{Vasiliev,Zinn,turbo,Tauber}.

In work \cite{AGKL17_pre} the velocity part (without $\beta_w$) of the system
 (\ref{eq:all_beta}) was analyzed.
 Altogether three IR attractive fixed points, which defines possible scaling regimes of the system, were 
 found.
The fixed point \fp{I} (the trivial or Gaussian point) is stable if $y$, $\varepsilon<0$. The coordinates are
\begin{equation}
  g_1^* = 0, \quad g_2^* = 0.
  \label{fp1}
\end{equation}
The fixed point \fp{II}, which is stable if $\varepsilon>0$ and $y<3\varepsilon/2$, has the following coordinates
\begin{equation}
  g_1^* = 0, \quad g_2^* = \frac{8\eps}{3}.
  \label{eq:fp2}
\end{equation}
The fixed point \fp{III} (stable if $y>0$ and $y>3\varepsilon/2$) is
\begin{equation}
  g_1^* = \frac{16y(2y-3\eps)}{9[y(2+\alpha)-3\eps]}, 
  \quad g_2^* =\frac{16\alpha y^2}{9[y(2+\alpha)-3\eps]}.
  \label{eq:fp3}
\end{equation}
The crossover between the two nontrivial points~(\ref{eq:fp2}) and~(\ref{eq:fp3}) takes place across the
line $y=3\eps/2$, which is in accordance with results of~\cite{Ant04}.

Moreover, from the analysis in \cite{AGKL17_pre} it follows that for nontrivial regimes
the coordinate $u$ takes value $u^*=1$. Substituting these values
together with $d=4$ we obtain for the charge $w$ the following beta function
\begin{equation}
   \beta_w = \frac{w-1}{16(1+w)^2}   
   \biggl[ 
   g_1(6+2\alpha+9w+3w^2)+g_2(3w^2+9w+8)
   \biggl].   
   \label{eq:betaw}
\end{equation}
Note that this result is in accordance with previous work for the passive scalar case \cite{AGKL17_epj}
and vector case as well \cite{AHHJ03}.
The expression in the square brackets in Eq. (\ref{eq:betaw}) is always positive
for physically permissible values, i.e., $g_1 > 0,g_2>0,w>0$ and $\alpha>0$. Therefore, only
one nontrivial solution for the fixed point exists, $w^*=1$. Also it is rather
straightforward to show that $\partial_w \beta_w >0$ at nontrivial fixed points, what
ensures IR stability.

Depending on the values of $y$ and  $\varepsilon$, the different values of the critical 
dimension for
various quantities $F$ are obtained. They can be calculated via the expression
\begin{equation}
  \Delta[F]=d^{k}_{F}+ \Delta_{\omega}d^{\omega}_{F} + \gamma_{F}^{*},
  \label{eq:Delta}
\end{equation}
where $d^{\omega}_{F}$ is the canonical frequency dimension, $d^{k}_{F}$ is the momentum dimension,
$\gamma_F^*$ is the anomalous dimension at the critical point (FPII or FPIII), and
$\Delta_{\omega}=2-\gamma_\nu^*$ is the critical dimension of frequency.

{Using Eq.~(\ref{eq:Delta}) the critical dimension} of the passive 
scalar density field $\theta$ 
and the field $\theta'$ were obtained for the fixed points \fp{II} and \fp{III}:
\begin{align}
  \Delta_{\theta_i} & = -1+\varepsilon/4, \quad \Delta_{\theta_i'}= d+1 -\varepsilon/4 \quad \text{for the fixed 
  point \fp{II}};
  \label{eq:KriTet1} \\
  \Delta_{\theta_i} & = -1+y/6, \quad \Delta_{\theta_i'}= d+1 -y/6  \quad \text{for the fixed point \fp{III}}.
  \label{eq:KriTet2}
\end{align}

{Measurable quantities are} some correlation functions or structure functions of composite operators.
A local composite operator is a monomial or polynomial constructed from the primary fields
$\theta(x)$ and their finite-order derivatives at a single space-time point $x$. In the Green functions with such
objects, new UV divergences arise due to the coincidence of
the field arguments. They can be removed by the additional
renormalization procedure \cite{Vasiliev,Collins}. 

The simplest case of a composite operator is the scalar operator $F(x)=\theta^n(x)$.  Here, we focus on the
irreducible tensor operators {of the form}
\begin{equation}
  F^{(n,l)}_{i_{1}\dots i_{l}} =
  \theta_{i_1}\cdots \theta_{i_l}\,
  ( \theta_j \theta_j)^{s} + \dots,
  \label{eq:Fnp}
\end{equation}
where $l$ is the number of the free vector indices (the rank of the tensor) and $n=l+2s$ is the total number 
of the fields $\theta$ entering the
operator. The ellipsis stands for the subtractions with the Kronecker's delta symbols that make the operator
irreducible (so that a contraction
with respect to any pair of the free tensor indices vanish). For instance,
\begin{equation}
  F^{(2,2)}_{ij} = \theta_i \theta_j -
  \frac{\delta_{ij}}{d}\, (\theta_k \theta_k).
  \label{eq:Irr}
\end{equation}

For practical calculations, it is convenient to contract the tensors
(\ref{eq:Fnp}) with an arbitrary constant vector
{\mbox{\boldmath $\lambda$}}$=\{\lambda_{i}\}$.
The resulting scalar operator {takes the form}
\begin{equation}
  F^{(n,l)} = (\lambda_{i}w_{i})^{l} (w_{i}w_{i})^{s} + \dots,
  \quad w_{i} \equiv \partial_{i}\theta,
  \label{eq:FnpSk}
\end{equation}
where the subtractions, denoted by the ellipsis, necessarily include the
factors of $\lambda^{2}=\lambda_{i}\lambda_{i}$. 

In order to calculate the critical dimension of the operator, one has to renormalize it. The 
operators~(\ref{eq:Fnp}) can be treated as
multiplicatively renormalizable, $F^{(n,l)} = Z_{(n,l)} F^{(n,l)}_{R}$, with certain renormalization
constants $Z_{(n,l)}$ (see \cite{AK14}). 
The counterterm to $F^{(n,l)}$ must have the same rank as the operator itself. It means that the terms 
containing $\lambda^{2}$  should be
excluded since the contracted fields $w_{i}w_{i}$, standing near them, reduce the number of free indices.
It is sufficient to retain only 
the principal monomial, explicitly shown in~(\ref{eq:FnpSk}),
and to discard in the result all the terms with factors of $\lambda^{2}$.
The renormalization constants $Z_{(n,l)}$ are determined by the finiteness of the 1-irreducible Green 
function $\Gamma_{nl}(x; \theta)$, which 
in the one-loop approximation is diagrammatically {represented as}
\begin{equation}
  \Gamma_{nl}(x; \theta) = F^{(n,l)} + \frac{1}{2}   
   \raisebox{-4.ex}{ \includegraphics[width=1.2truecm]{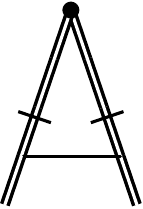}},
  \label{eq:dom}
\end{equation}
where numerical factor $1/2$ is a symmetry factor of the graph and the thick dot with two lines attached
denotes the operator vertex
\begin{equation}
  V(x;x_{1},x_{2})=  \frac{\delta^{2}F^{(n,l)}}{\delta \theta_i \delta \theta_j}.
  \label{eq:Vae1}
\end{equation}
Divergent parf of a one-loop diagram in (Eq.~\ref{eq:dom}) reads
\begin{align*}   
  \raisebox{-4.ex}{ \includegraphics[width=1.2truecm]{comp.pdf}}
  & = -\frac{F_{nl}}{2wd(d+2)}\biggl[  
  \frac{ Q_1}{(1+w)} \biggl( \frac{g_1}{y} + \frac{g_2}{\eps} \biggl)
  +
  \frac{ Q_2}{u(u+w)} \biggl( \frac{\alpha g_1}{y} + \frac{g_2}{\eps} \biggl)
  \biggl]
  ;\\
  Q_1 &\equiv  -n(n+d)(d-1)-l(d+1)(d+l-2) ,\\
  Q_2 & \equiv  -n(d-1)(nd+n-d)-l(d+l-2).
\end{align*}

The expressions for the propagators and vertices at the bottom of the diagram can be 
found in~\cite{AGKL17_pre}. 
Then using the chain rule and up to irrelevant terms the vertex~(\ref{eq:Vae1}) for the operator
$F^{(n,l)}$ can be presented in the form 
\begin{equation}
  V(x;x_{1},x_{2})=  \frac{\partial^{2}F^{(n,l)}}{\partial w_{i}\partial w_{j}}
  \, \delta(x-x_{1})\, \delta(x-x_{2}).
  \label{eq:Vae10}
\end{equation}
The differentiation yields
\begin{align}
  {\partial^{2}F^{(n,l)}}/{\partial w_{i}\partial w_{j}} & = 
  2s (w^{2})^{s-2} (\lambda w)^{l} \left[\delta_{ij} w^{2} +2(s-1)w_{i}w_{j}
  \right] + l(l-1) (w^{2})^{s} \nonumber \\
  &\times (\lambda w)^{l-2} \lambda_{i} \lambda_{j}+
   +  2ls (w^{2})^{s-1}(\lambda w)^{l-1} (w_{i}\lambda_{j}+ w_{j}\lambda_{i}),
  \label{eq:Vae11}
\end{align}
where $w^{2}=w_{k}w_{k}$, $(\lambda w)=\lambda_{k}w_{k}$
and substitution $w_i \rightarrow \theta_i$ is assumed.
 Two more
factors $w_{p}w_{r}$ are attached to the bottom of the diagram due to the
derivatives coming from the vertices $\theta'_i(v_k\partial_k)\theta_i$. The ultraviolet
divergence is logarithmic and one can set all the external frequencies and
momenta equal to zero; then the core of the diagram takes the form
\begin{equation}
  \int\frac{\text{d}\omega}{2\pi}
  \int_{k>m}\frac{{\dRM^d}{k}}{(2\pi)^{d}}\,
  k_{i}k_{j}\, D_{pr} (\omega, {\mk})\,
  \frac{1}{\omega^{2}+w^{2}\nu^2 k^{4}}.
  \label{eq:triad}
\end{equation}
Here the first factor comes from the derivatives in~(\ref{eq:Vae1}), $w=\kappa/\nu$, $D_{pr}$ is the velocity
correlation function [see~(\ref{eq:correl2})], and
the last factor comes from the two propagators $\langle\theta'\theta\rangle_{0}$.

After the integration, combining all the
factors, contracting the tensor indices and expressing the result in terms of $n=l+2s$ and $l$, one obtains: 
\begin{align}
  \Gamma_{n}(x;\theta) & = F^{(n,l)}(x)\, \biggl\{ 1 - \frac{1}{4wd(d+2)}\,
  \biggl[
  \frac{Q_{1}}{(1+w)} \biggl( \frac{g_1 }{y} + \frac{g_2 }{\eps}\biggl) \nonumber\\
  & + 
  \frac{Q_{2}}{u(u+w)}\biggl( \frac{\alpha g_1}{y} +  \frac{ g_2 }{\eps}\biggl)
  \biggl] \biggl\}.
  \label{eq:QQ}
\end{align}
Then the renormalization constants $Z_{(n,l)}$ calculated in the MS scheme read
\begin{equation}
  Z_{(n,l)} = 1 - \frac{1}{4dw(d+2)}\,
  \left[
  \frac{Q_{1}}{1+w} \left(\frac{ g_{1} }{y}+\frac{ g_{2} }{\eps}\right) + 
  \frac{Q_{2}}{u(u+w)}\left( \frac{\alpha g_{1} }{y}+\frac{ g_{2} }{\eps}\right)
  \right].
  \label{eq:ZZ}
\end{equation}
For the corresponding anomalous dimension one obtains
\begin{equation}
  \gamma_{(n,l)} = \frac{1}{4dw(d+2)}\, \left\{
  \frac{Q_{1}}{1+w} (g_{1}+g_{2}) +  \frac{Q_{2}}{u(u+w)} (\alpha g_{1}+g_{2})
  \right\}.
  \label{eq:GG}
\end{equation}

In order to evaluate the critical dimension, one needs to substitute the coordinates of the fixed 
points into the  
expression~(\ref{eq:GG}) and then use the relation~(\ref{eq:Delta}). For the fixed point FPII the critical dimension is
\begin{equation}
  %\Delta_{(n,l)} = \frac{n}{4}\varepsilon + \frac{Q_1+Q_2}{3d(d+2)}\varepsilon.
  \Delta_{(n,l)} = \frac{n}{4}\varepsilon + \frac{Q_1+Q_2}{72}\varepsilon.
  \label{eq:Dnl1}
\end{equation}
For the fixed point FPIII it is
\begin{equation}
  \Delta_{(n,l)} = \frac{n}{6}y + \frac{y}{12}\frac{Q_1 (\alpha y + 2y - 3\varepsilon) + 3\alpha Q_2 (y-\varepsilon)}{9[y(2+\alpha)-3\varepsilon]}.
  \label{eq:Dnl2}
\end{equation}
Both expressions~\eqref{eq:Dnl1} and~\eqref{eq:Dnl2} suppose higher order corrections in $y$ and 
$\varepsilon$. 

Therefore, the infinite set of  operators with negative critical dimensions, whose spectra is unbounded
from below, is observed.

%---------------------------------------------------------------------------------------------------------------
\section{Operator Product Expansion}\label{sec:ope}
%---------------------------------------------------------------------------------------------------------------
Our main interest are pair correlation functions, whose
unrenormalized counterparts have been defined in Eq. (\ref{eq:Fnp}).
For Galilean invariant equal-time functions we can write
 the following representation
\begin{equation}
  \langle F^{(m,i)}(t,\mx) F^{(n,j)} (t,\mx') \rangle \simeq \mu^{d_F} \nu^{d^\omega_F} (\mu r)^{-\Delta_{(m,i)}-
  \Delta_{(n,j)}} \zeta_{m,i;n,j}(mr, c(r)),  
  \label{eq:struc}
\end{equation}
where $r=|\mx-\mx'|$ and $c(r)$ is effective speed of sound. Its limiting behaviour
can be shown  \cite{AGKL17_pre} to be 
\begin{equation*}
  c(r) = c \frac{(\mu r)^{\Delta_c}  }{\mu\nu} \rightarrow
         \begin{cases}
           c(0) \quad \text{{for} non-local r;} \\
           c(\infty) \quad \text{{for} local r.} 
         \end{cases}
\end{equation*}
 Eq.~(\ref{eq:struc}) is valid  in the
asymptotic limit $\mu r\gg 1$. Further, the inertial-convective range corresponds to the additional
restriction $mr\ll 1$. The behaviour of the
functions $\zeta$ at $mr\to0$ can be studied by means of the OPE technique \cite{Vasiliev,Collins}.
The basic idea of this method is to represent a product of two operators
at two close points, $\mx$ and $\mx'$ with $\mx-\mx'\rightarrow 0$, in the form
\begin{equation}
  F^{(m,i)}(t,\mx)F^{(n,j)}(t,\mx') \simeq \sum_{F} C_{F}(mr)\,
  F \biggl(t,\frac{\mx + \mx'}{2} \biggl),
  \label{eq:OPE2}
\end{equation}
where functions $C_F$ are regular in their argument and a given sum runs over
all permissible local composite operators $F$ allowed by RG and symmetry considerations.
 Taken into account (\ref{eq:struc}) and (\ref{eq:OPE2}) in the limit $mr \rightarrow 0$  we arrive
 at the relation
	\begin{equation*}
	   \zeta(mr) \approx \sum_F  A_F(mr) 
	   (mr)^{\Delta_F}
	\end{equation*}
Considering OPE for the correlation functions $\langle F^{(p,0)} F^{(k,0)} \rangle$ with $n=p+k$,
 where 
$F^{(n,l)}$ is the operator of the type~(\ref{eq:Fnp}), one
can observe that the leading contribution to the expansion is determined by the
operator $F^{(n,0)}$ from 
the same family. Therefore, 
in the inertial range these correlation functions acquire the form
\begin{equation}
  \langle F^{(p,0)}(t,{\mx}) F^{(k,0)}(t,{\mx'}) \rangle \sim
  r^{-\Delta_{(p,0)}-\Delta_{(k,0)}+\Delta_{(n,0)}}.
  \label{eq:MF}
\end{equation}
The inequality $\Delta_{(n,0)}<\Delta_{(p,0)}+\Delta_{(k,0)}$, which follows from both explicit one-loop
expressions~(\ref{eq:Dnl1}) and~(\ref{eq:Dnl2}), indicates,
that the operators $F^{(n,0)}$ demonstrate a ``multifractal'' behaviour; see \cite{DL}.

A direct substitution of $d=4$ leads to the following prediction for a critical dimension
\begin{equation}
   \Delta_{n,l} = n\Delta_\theta + \gamma^*_{nl} = 
   \begin{cases}
      -n + \frac{n\eps}{4} +\frac{(Q_1+Q_2)\eps}{4} \qquad \text{for FPII} \\
      -n + \frac{n y}{6} + \frac{Q_1 y}{108} + \frac{Q_2 \alpha y(y-\eps)}{36[y(2+\alpha)-3\eps]}
      \qquad \text{for FPIII}
   \end{cases}
   \label{eq:OPE_res}
\end{equation}
where we have
\begin{equation*}
  Q_1|_{d=4} = -3n(n+4)+ 5l(2+l) ,\quad 
  Q_2|_{d=4} = -3n(5n-4)+l(2+l).
\end{equation*}
From these results several observations can be made.
 Based on (\ref{eq:OPE_res}) we see that for fixed $n$ kind of a hierarchy present with respect to
the index $l$, i.e,
        \begin{equation}
            \frac{\partial \Delta_{n,l} }{\partial l} >0.
        \end{equation}
 In other words, the higher $l$ the less important contribution. The most relevant
 is given by the isotropic shell with $l=0$. This is in accordance with
 previous studies \cite{AHHJ03,AK14,AK15}. Moreover, we observe that there is
 no appearance of the parameter $\alpha$ for the local regime FPII.
  Further, in contrast to \cite{AK15} there is no monotonous behaviour in $\alpha$ of
        $\Delta_{n0}$ for the non-local regime.
%---------------------------------------------------------------------------------------------------------------
\section{Conclusion}\label{sec:conclusion}
%---------------------------------------------------------------------------------------------------------------

In the present paper the advection of the vector field by the Navier-Stokes velocity
ensemble has been examined. The fluid was assumed to be
compressible and the space dimension was close to $d=4$. The problem has been investigated by means of 
renormalization group and operator product expansion; 
the double expansion in $y$ and $\varepsilon=4-d$ was constructed. 

There are two nontrivial IR stable fixed points in this model and, therefore, the critical behaviour in the inertial 
range demonstrates two different regimes depending on the 
relation between the exponents $y$ and $\varepsilon$. The expressions for the critical exponents of the
{vector} field $\theta$ were obtained in the leading one-loop approximation. 

In order to find the anomalous exponents of the structure functions, the composite fields~(\ref{eq:Fnp}) were 
renormalized. The critical dimensions of them were evaluated. 
It turned out that there is an infinite number of the dangerous operators, i.e., the operators with negative
critical dimensions. Besides, OPE allowed us to derive the 
explicit expressions for the critical dimensions of the structure functions. 
The existence of the anomalous scaling in the inertial-convective range was established for both possible
scaling regimes.
Another very interesting result is that some kinds of operators exhibit the ``multifractal'' behaviour.

With regard to future research, it would be interesting to go beyond the one-loop approximation and to 
analyze the behaviour more precisely on the higher level of accuracy.
{Another very important task to be further 
investigated is to have a closer look at the both scalar and vector active fields, 
i.e., to consider a back influence of the advected fields to the turbulent environment flow.}

%------------------------------------------------------------------------------------------------%
\section*{Acknowledgements}
The work was supported by VEGA grant No.~1/0345/17 of the Ministry
of Education, Science, Research and Sport of the Slovak Republic,
by the Ministry of Education and Science of Russian Federation
(the Agreement number 02.a03.21.0008), {and  by the Russian Foundation
for Basic Research within the Project No. 16-32-00086}.
N.~M.~G. acknowledges the support from the Saint Petersburg Committee of Science and High School.

%------------------------------------------------------------------------------------------------%

\end{document}